\documentclass[12pt,thmsa]{article}
\usepackage{amssymb}

\input tcilatex
\QQQ{Language}{American English}

\begin{document}

\begin{center}
\emph{The derivation of the coupling constant in the new Self Creation
Cosmology}

Garth A Barber

The Vicarage, Woodland Way, Tadworth, Surrey, England KT206NW

Tel: +44 01737 832164 \qquad

e-mail: garth.barber@virgin.net

\allowbreak \textit{Abstract}
\end{center}

It has been shown that the new Self Creation Cosmology theory predicts a universe with
a total density parameter of one third yet spatially flat, which would
appear to accelerate in its expansion. Although requiring a moderate amount
of 'cold dark matter' the theory does not have to invoke the hypotheses of
inflation, 'dark energy', 'quintessence' or a cosmological constant
(dynamical or otherwise) to explain observed cosmological features. The
theory also offers an explanation for the observed anomalous Pioneer
spacecraft acceleration, an observed spin-up of the Earth and an problematic
variation of G observed from analysis of the evolution of planetary
longitudes. It predicts identical results as General Relativity in standard
experimental tests but three definitive experiments do exist to falsify the
theory. In order to match the predictions of General Relativity, and
observations in the standard tests, the new theory requires
the Brans Dicke omega parameter that couples the scalar field to matter to
be -3/2 . Here it is shown how this value for the coupling parameter is
determined by the theory's basic assumptions and therefore it is an inherent
property of the principles upon which the theory is based.

\begin{center}
\ 
\end{center}

\section{Introduction}

\subsection{Notations and conventions}

This paper adopts the ''Landau-Lifshitz Spacelike Convention''. A comma
denotes ordinary differentiation, a semi-colon denotes covariant
differentiation, and $\Box \phi $ is the d'Alembertian invariant, $\phi
_{;\;;\sigma }^\sigma $ . $\mathbf{\nabla }$ is the normal gradient. The
number of dimensions of the manifold $M$ is four throughout, the summation
convention is followed, Greek indices indicate four-space-time and Latin
indices indicate three-space. The metric signature is ($-,+,+,+$), the Ricci
tensor is given in terms of the Christoffel symbols $\Gamma _{\;\nu
\varepsilon }^\mu $ by 
\[
+R_{\mu \nu }=+\Gamma _{\;\mu \nu ,\rho }^\rho -\Gamma _{\;\rho \nu ,\mu
}^\rho +\Gamma _{\;\mu \nu }^\sigma \Gamma _{\;\sigma \rho }^\rho -\Gamma
_{\;\rho \nu }^\sigma \Gamma _{\;\sigma \mu }^\rho 
\]
$G_N$ is the Newtonian gravitational constant as measured in ''Cavendish''
type experiments and the speed of light is unity unless otherwise specified
as $c$. A tilde signifies the Einstein frame.

\subsection{An historical introduction to Self Creation Cosmology}

In a recent paper this author, [Barber,(2002a)], developed a new Self
Creation Cosmology (SCC), which superceeded and subsumed two earlier versions
[Barber, (1982)]. All versions of SCC were modifications of the Brans Dicke
theory (BD), [Brans \& Dicke, (1961)] in which the equivalence principle was
relaxed to allow for the continuous creation of mass out of self contained
matter, gravitational and scalar fields. The new theory is a 'semi-metric' theory
in that photons follow trajectories that are geodesics of the theory but particles 
do not, \textit{in\ vacuo} they follow orbits that are identical to General Relativity 
geodesics. The new theory was shown to predict a universe with a total density parameter 
of one third yet spatially flat, although it did require a moderate amount of 'cold dark 
matter'. It did not have to invoke the hypotheses of inflation, 'dark energy',
'quintessence' or a cosmological constant (dynamical or otherwise) to
explain observed cosmological features. The new theory also predicted
identical results as General Relativity (GR) in standard experimental tests.
In a subsequent paper, [Barber, (2002b)], it was shown how the theory also
offered an explanation for 'cosmic acceleration', the anomalous Pioneer
spacecraft acceleration, an observed spin-up of the Earth and an apparent variation of
G observed from analysis of the evolution of planetary longitudes. Finally,
it was shown, [Barber, (2003)], that measurement of geodetic precession, about to be 
performed on the Gravity Probe B satellite is a definitive experiment, which 
could falsify the theory as SCC predicts a value $\frac 23$ of that expected
by GR.

For SCC to agree with observation, and hence GR, in the standard tests a
coupling parameter $\omega $ was empirically set with the value 
\begin{equation}
\omega =-\frac 32\text{ .}  \label{eq1a}
\end{equation}
This paper will show why is value of $\omega $ has not been arbitrarily
chosen but it is a definitive value required for consistency by the basic
principles of the theory.

\subsection{A summary of the theory}

Mach's Principle (MP) is incorporated in SCC by assuming the inertial masses
of fundamental particles are dependent upon their interaction with a scalar
field $\phi $ coupled to the large scale distribution of matter in motion in
a similar fashion as BD. This coupling is described by a field equation of
the simplest general covariant form 
\begin{equation}
\Box \phi =4\pi \lambda T_{M\;\sigma }^{\;\;\sigma }\text{ ,}  \label{eq2}
\end{equation}
$T_{M\;\sigma }^{\;\;\sigma }$ is the trace of the energy momentum tensor
describing all non-gravitational and non-scalar field energy. The BD
coupling parameter $\lambda $ was found to be unity. [Barber, (2002a)] and
in the spherically symmetric One Body problem 
\begin{equation}
\stackunder{r\rightarrow \infty }{Lim}\phi \left( r\right) =\frac \psi {G_N}%
\text{ ,}  \label{eq1}
\end{equation}
where $G_N$ is the normal gravitational constant measured in Cavendish type
experiments and $\psi $ is a constant. $\psi $ was found to be unity in the
subsequent development of the theory as described below.

In BD the Equivalence Principle was retained so that particle rest masses
were invariant and $G$ would be observed to vary with position. By contrast,
in SCC in which the Equivalence Principle is violated, it is $G_N$ that is measured
to be invariant and it is particle rest masses that vary.

In all the SCC theories the Equivalence Principle is relaxed, specifically
in the new theory it is replaced by the Principle of Mutual Interaction
(PMI) in which 
\begin{equation}
T_{M\;\nu ;\mu }^{.\;\mu }=f_\nu \left( \phi \right) \Box \phi =4\pi f_\nu
\left( \phi \right) T_{M\;\sigma }^{\;\;\sigma }\text{ .}  \label{eq7}
\end{equation}
Therefore \textit{in\ vacuo,} 
\begin{equation}
T_{em\,\nu ;\mu }^{\quad \mu }=4\pi f_\nu \left( \phi \right) T_{em\;\sigma
}^{\;\;\sigma }=4\pi f_\nu \left( \phi \right) \left( 3p_{em}-\rho
_{em}\right) =0  \label{eq8}
\end{equation}
where $p_{em}$ and $\rho _{em}$ are the pressure and density of an
electromagnetic radiation field with an energy momentum tensor $T_{em\,\mu
\nu }$ and where $p_{em}=\frac 13\rho _{em}$ . Thus the scalar field is a
source for the matter-energy field if and only if the matter-energy field is
a source for the scalar field. Although the equivalence principle is
violated for particles, it is not for photons, which still travel through
empty space on (null) geodesic paths.

Particles do not have invariant rest mass and its variation is determined by
a second principle, the Local Conservation of Energy. This requires the
energy used in lifting a particle against a gravitational field to be
absorbed into its rest mass; so that rest masses include gravitational
potential energy. A particle's rest mass is described by 
\begin{equation}
m_p(x^\mu )=m_0\exp [\Phi _N\left( x^\mu \right) ]\text{ ,}  \label{eq24}
\end{equation}
where $\Phi _N\left( x^\mu \right) $ is the dimensionless Newtonian
potential and $m_p\left( r\right) \rightarrow m_0$ as $r\rightarrow \infty $
.

In SCC gravitational orbits and cosmological evolution are described in the
Jordan frame. A conformal equivalence exits between the Jordan frame and
canonical GR, which is the theory's Einstein frame. This results in the 
geodesic orbits of SCC being identical with GR \textit{in vacuo}. The Jordan 
(energy) frame [JF(E)] conserves mass-energy, and the Einstein frame (EF) conserves 
energy momentum. The two conformal frames are related by a coordinate transformation 
\begin{equation}
g_{\mu \nu }\rightarrow \widetilde{g}_{\mu \nu }=\Omega ^2g_{\mu \nu }\text{
.}  \label{eq9}
\end{equation}

The JF of SCC requires mass creation, ( $T_{M\;\nu ;\mu }^{\;\mu }\neq 0$ ),
therefore the scalar field is non-minimally connected to matter. The JF
Lagrangian density is, 
\begin{equation}
L^{SCC}[g,\phi ]=\frac{\sqrt{-g}}{16\pi }\left( \phi R-\frac \omega \phi
\phi _{;\sigma }\phi _{;}^\sigma \right) +L_{matter}^{SCC}[g,\phi ]\text{ ,}
\label{eq24a}
\end{equation}
and its conformal dual, [Dicke (1962)], by a general transformation $%
\widetilde{g}_{\mu \nu }=\Omega ^2g_{\mu \nu }$ , is 
\begin{eqnarray}
L^{SCC}[\widetilde{g},\widetilde{\phi }] &=&\frac{\sqrt{-\widetilde{g}}}{%
16\pi }\left[ \widetilde{\phi }\widetilde{R}+6\widetilde{\phi }\widetilde{%
\Box }\ln \Omega \right] +\widetilde{L}_{matter}^{SCC}[\widetilde{g},%
\widetilde{\phi }]  \label{eq24b} \\
&&\ \ \ -\frac{\sqrt{-\widetilde{g}}}{16\pi }[\left( 2\omega +3\right) \frac{%
\Omega _{;\sigma }\Omega _{;}^\sigma }{\Omega ^2}  \nonumber \\
&&\ \ \ +4\omega \frac{\Omega _{;\sigma }\widetilde{\phi }_{;}^\sigma }%
\Omega +\omega \frac{\widetilde{\phi }_{;\sigma }\widetilde{\phi _{;}^\sigma 
}}{\widetilde{\phi }}]\text{ .}  \nonumber
\end{eqnarray}

A mass is conformally transformed according to 
\begin{equation}
m\left( x^\mu \right) =\Omega \widetilde{m}_0\text{ ,}  \label{eq24c}
\end{equation}
Equation \ref{eq24} requires 
\begin{equation}
\Omega =\exp \left[ \Phi _N\left( x^\mu \right) \right] \text{ ,}
\label{eq34}
\end{equation}
where $m\left( x^\mu \right) $ is the mass of a fundamental particle in the
JF and $\widetilde{m}_0$ its mass in the EF.

If we define the EF by $G$ $=\,G_N$ a constant, i.e. $\widetilde{\phi }%
_{;\sigma }=0$ , and as \textit{in vacuo } 
\begin{equation}
\widetilde{\Box }\ln \Omega =\widetilde{\Box }\Phi _N\left( x^\mu \right)
=\nabla ^2\Phi _N\left( x^\mu \right) =0\text{ , }  \label{eq24e}
\end{equation}
setting $\omega =-\frac 32$ reduces the EF conformal dual, the EF Lagrangian
density to 
\begin{equation}
L^{SCC}[\widetilde{g},\widetilde{\phi }]=\frac{\sqrt{-\widetilde{g}}}{16\pi }%
\left[ \widetilde{\phi }\widetilde{R}\right] +\widetilde{L}_{matter}^{SCC}[%
\widetilde{g},\widetilde{\phi }]\text{ ,}  \label{eq24d}
\end{equation}
which is canonical GR.

The question is, ''Why should $\omega =-\frac 32$ ?'' It is the purpose of the present paper 
to answer this question.

\section{Incorporating the Principle of Mutual Interaction in BD}

\subsection{ Constructing the SCC Field Equations}

The first basis of the new Self Creation Cosmology is to fully incorporate
Mach's Principle in GR by following BD but then introducing the PMI,
Equation \ref{eq7}, to quantify the violation of the Equivalence Principle
and to allow for mass creation.

A scalar field $\phi $ is defined by Equations \ref{eq2} and \ref{eq1}. An
energy momentum tensor for the scalar field $T_{\phi \,\,\mu \nu }$ is added
to the gravitational field equation in order to account for its presence and
effect on the curvature of space-time 
\begin{equation}
R_{\mu \nu }-\frac 12g_{\mu \nu }R=\frac{8\pi }\phi \left( T_{M\;\mu \nu
}+T_{\phi \;\mu \nu }\right) \text{ .}  \label{eq3}
\end{equation}
In order to determine $T_{\phi \,\mu \nu }$ and $f_\nu \left( \phi \right) $
Equation \ref{eq3} is written in the following mixed tensor form 
\begin{equation}
T_{M\;\nu }^{\;\mu }=\frac \phi {8\pi }\left( R_{\;\nu }^\mu -\frac 12\delta
_{\;\nu }^\mu R\right) -T_{\phi \;\nu }^{\;\mu }  \label{eq5}
\end{equation}
and the method described by Weinberg, [Weinberg, (1972), pages 158-160, equations 7.3.4-7.3.12]
 is followed. 
The most general form of $T_{\phi \;\nu }^{\;\mu }$ using two derivatives of one
or two $\phi $ fields and $\phi $ itself is 
\begin{equation}
T_{\phi \;\nu }^{\;\mu }=A(\phi )\phi ;^\mu \phi ;_\nu +B(\phi )\delta
_{\;\nu }^\mu \phi ;_\sigma \phi ;^\sigma +C(\phi )\phi ;^\mu ;_\nu +D(\phi
)\delta _{\;\nu }^\mu \Box \phi  \label{eq12}
\end{equation}
and covariantly differentiating produces 
\begin{eqnarray}
T_{\phi \;\nu ;\mu }^{\;\mu } &=&\left[ A^{\prime }(\phi )+B^{\prime }(\phi
)\right] \phi ;^\mu \phi ;_\nu \phi ;_\mu +\left[ A(\phi )+D^{\prime }(\phi
)\right] \phi ;_\nu \Box \phi  \label{eq13} \\
&&\ \ +\left[ A(\phi )+2B(\phi )+C^{\prime }(\phi )\right] \phi ;^\mu ;_\nu
\phi ;_\mu  \nonumber \\
&&\ \ +D(\phi )\left( \Box \phi \right) ;_\nu +C(\phi )\Box (\phi ;_\nu )%
\text{ ,}  \nonumber
\end{eqnarray}

where a prime ($^{\prime }$) means differentiation w.r.t $\phi $. Now make
use of the Bianchi identities and the identity (observing the sign
convention) 
\begin{equation}
\phi ;_\sigma R_{\;\nu }^\sigma =\Box \left( \phi ;_\nu \right) -\left( \Box
\phi \right) ;_\nu  \label{eq14}
\end{equation}
in order to examine the violation of the equivalence principle. Covariantly
differentiating Equation \ref{eq5} yields 
\begin{equation}
T_{M\;\nu ;\mu }^{\;\mu }=\frac{\phi ;_\mu }{8\pi }\left( R_{\;\nu }^\mu
-\frac 12\delta _{\;\nu }^\mu R\right) -T_{\phi \;\nu ;\mu }^{\;\mu }\text{ .%
}  \label{eq15}
\end{equation}
Taking the trace of Equation \ref{eq3} gives 
\begin{equation}
R=-\frac{8\pi }\phi \left[ T_{M\;\sigma }^{\;\sigma }+T_{\phi \;\sigma
}^{\;\sigma }\right] \text{ ,}  \label{eq16}
\end{equation}
with 
\begin{equation}
T_{\phi \;\sigma }^{\;\sigma }=\left[ A(\phi )+4B(\phi )\right] \phi
;^\sigma \phi ;_\sigma +\left[ C(\phi )+4D(\phi )\right] \Box \phi \text{ ,}
\label{eq17}
\end{equation}
and from Equation \ref{eq2} 
\begin{equation}
T_{M\;\sigma }^{\;\sigma }=\frac 1{4\pi \lambda }\Box \phi \text{ .}
\label{eq18}
\end{equation}
Substituting Equations \ref{eq17} and \ref{eq18} in Equation \ref{eq16}
yields 
\begin{equation}
R=-\frac{8\pi }\phi \left\{ \left[ A(\phi )+4B(\phi )\right] \phi ;^\sigma
\phi ;_\sigma +\left[ C(\phi )+4D(\phi )+\frac 1{4\pi \lambda }\right] \Box
\phi \right\} \text{ .}  \label{eq19}
\end{equation}
While Equations \ref{eq13}, \ref{eq14} and \ref{eq19} substituted in \ref
{eq15} produce 
\begin{eqnarray}
T_{M\;\nu ;\mu }^{\;\mu } &=&-\frac 1{8\pi }(\Box \phi );_\nu +\frac 1{8\pi
}\Box (\phi ;_\nu )+\frac 1{2\phi }\left[ A(\phi )+4B(\phi )\right] \phi
;^\mu \phi ;_\mu \phi ;_\nu  \label{eq20} \\
&&\ +\frac 1{2\phi }\left[ C(\phi )+4D(\phi )+\frac 1{4\pi \lambda }\right]
\phi ;_\nu \Box \phi  \nonumber \\
&&\ -\left[ A^{\prime }(\phi )+B^{\prime }(\phi )\right] \phi ;^\mu \phi
;_\nu \phi ;_\mu -\left[ A(\phi )+D^{\prime }(\phi )\right] \phi ;_\nu \Box
\phi  \nonumber \\
&&\ -\left[ A(\phi )+2B(\phi )+C^{\prime }(\phi )\right] \phi ;^\mu ;_\nu
\phi ;_\mu -D(\phi )\left( \Box \phi \right) ;_\nu -C(\phi )\Box (\phi ;_\nu
)\text{ .}  \nonumber
\end{eqnarray}
If the Principle of Mutual Interaction, Equation \ref{eq7}, is applied 
\[
T_{M\quad \nu ;\mu }^{.\quad \mu }=f_\nu \left( \phi \right) \Box \phi 
\]
So the coefficients of:$(\Box \phi );_\nu $; $\Box (\phi ;_\nu )$, $\phi
;^\mu ;_\nu \phi ;_\mu $, and $\phi ;^\mu \phi ;_\mu \phi ;_\nu $, must
vanish in Equation \ref{eq20}, but those of $\phi ;_\nu \Box \phi $ must
satisfy Equation \ref{eq7}. This yields five equations to solve for the five
functions; $A(\phi )$, $B(\phi )$, $C(\phi )$, $D(\phi )$ and $f_\nu (\phi )$%
. 
\[
\begin{tabular}{llllll}
\underline{Term} &  & \underline{Coefficients ( = 0)} &  & \underline{%
Solution} &  \\ 
&  &  &  &  &  \\ 
$(\Box \phi );_\nu $ &  & $-\frac 1{8\pi }-D(\phi )=0$ &  & $D(\phi )=-\frac
1{8\pi }$ & (i) \\ 
&  &  &  &  &  \\ 
$\Box (\phi ;_\nu )$ &  & $+\frac 1{8\pi }-C(\phi )=0$ &  & $C(\phi )=+\frac
1{8\pi }$ & (ii) \\ 
&  &  &  &  &  \\ 
$\phi ;^\mu ;_\nu \phi ;_\mu $ &  & $A(\phi )+2B(\phi )+C^{\prime }(\phi )=0$
&  & $A(\phi )=-2B(\phi )$ & (iii) \\ 
&  &  &  &  &  \\ 
$\phi ;^\mu \phi ;_\mu \phi ;_\nu $ &  & $\frac 1{2\phi }\left[ A(\phi
)+4B(\phi )\right] $ &  &  &  \\ 
&  & $-\left[ A^{\prime }(\phi )+4B^{\prime }(\phi )\right] =0$ &  &  & (iv)
\end{tabular}
\]
Substituting Equation (iii) into (iv) 
\begin{equation}
\frac{B^{\prime }(\phi )}{B(\phi )}=-\frac 1\phi  \label{eq21}
\end{equation}
which has the solution 
\begin{equation}
B(\phi )=\frac k\phi \text{ ,}  \label{eq22}
\end{equation}
where $k$ is a constant, and therefore by Equation (iii) 
\begin{equation}
A(\phi )=-\frac{2k}\phi \text{ .}  \label{eq23}
\end{equation}
If $\kappa $ is now written as 
\[
k=-\frac \omega {16\pi } 
\]
the non-unique solution is obtained 
\begin{equation}
\begin{tabular}{llll}
& $A(\phi )=\frac \omega {8\pi \phi }$ &  & $B(\phi )=-\frac \omega {16\pi
\phi }$ \\ 
&  &  &  \\ 
& $C(\phi )=\frac 1{8\pi }$ &  & $D(\phi )=-\frac 1{8\pi }$%
\end{tabular}
\text{ .}  \label{eq24f}
\end{equation}

This solution looks the same as the BD solution except that $\omega $ is as
yet undetermined. A solution for $T_{M\;\nu ;\mu }^{\;\mu }$ is obtained by
substituting Equation \ref{eq24f} into Equation \ref{eq20} and examining the
coefficients of $\phi ;_\nu \Box \phi $. This fifth equation which will
determine $\omega $ is 
\begin{equation}
T_{M\;\nu ;\mu }^{\;\mu }=\left( \frac 1{16\pi \phi }-\frac 1{4\pi \phi
}+\frac 1{8\pi \lambda \phi }-\frac \omega {8\pi \phi }\right) \phi ;_\nu
\Box \phi \text{ ,}  \label{eq25}
\end{equation}
which can be written as 
\begin{equation}
T_{M\;\nu ;\mu }^{\;\mu }=\frac \kappa {8\pi }\frac{\phi ;_\nu }\phi \Box
\phi \text{ ,}  \label{eq26}
\end{equation}
so 
\begin{equation}
f_\nu \left( \phi \right) =\frac \kappa {8\pi }\frac{\phi ;_\nu }\phi
\label{eq26a}
\end{equation}
where 
\begin{equation}
\kappa =\frac 1\lambda -\frac 32-\omega \text{ .}  \label{eq27}
\end{equation}
$\kappa $ can be thought of as an undetermined ''creation coefficient''.
Note however that if $\kappa =0$, i.e. when Equation \ref{eq26} reduces to 
\[
T_{M\;\nu ;\mu }^{\;\mu }=0 
\]
which is the normal GR and BD conservation equation, then 
\begin{equation}
\omega =\varpi =\frac 1\lambda -\frac 32  \label{eq27a}
\end{equation}
where $\varpi $ is the standard BD parameter, and the BD field equations
have been recovered as to be expected.

The complete set of field equations are now:

1.\ The scalar field Equation \ref{eq2}.

2.\ The gravitational field equation 
\begin{eqnarray}
R_{\mu \nu }-\frac 12g_{\mu \nu }R &=&\frac{8\pi }\phi T_{M\mu \nu }+\frac
\omega {\phi ^2}\left( \phi ;_\mu \phi ;_\nu -\frac 12g_{\mu \nu }\phi
;_\sigma \phi ;^\sigma \right)  \label{eq28} \\
&&+\frac 1\phi \left( \phi ;_\mu ;_\nu -g_{\mu \nu }\Box \phi \right) \text{
,}  \nonumber
\end{eqnarray}
where $\omega =\frac 1\lambda -\frac 32-\kappa $ is a constant and $\lambda $
and $\kappa $ are coefficients yet to be determined.

3.\ The ''creation'' field Equation \ref{eq26} developed from the PMI, which
replaces the conservation equation in GR and BD.

The source of curvature $S_{\mu \nu }$ is defined by 
\begin{equation}
R_{\mu \nu }=\frac{8\pi }\phi S_{\mu \nu }  \label{eq29}
\end{equation}
where $S_{\mu \nu }$ is derived from Equation \ref{eq28} to be 
\begin{eqnarray}
S_{\mu \nu } &=&T_{M\,\mu \nu }-\frac 12\left( 1-\frac 12\lambda \right)
g_{\mu \nu }T_{M\;\sigma }^{\;\sigma }+\frac 1{8\pi \phi }\varpi \phi ;_\mu
\phi ;_\nu  \label{eq30} \\
&&\ +\frac 1{8\pi }\phi ;_\mu ;_\nu \text{ .}  \nonumber
\end{eqnarray}
The gravitational field equation can be written 
\begin{eqnarray}
R_{\mu \nu } &=&\frac{8\pi }\phi \left[ T_{M\,\mu \nu }-\frac 12\left(
1-\frac 12\lambda \right) g_{\mu \nu }T_{M\;\sigma }^{\;\sigma }\right]
+\frac \varpi {\phi ^2}\phi ;_\mu \phi ;_\nu  \label{eq31} \\
&&\ +\frac 1\phi \phi ;_\mu ;_\nu  \nonumber
\end{eqnarray}
so $R_{\mu \nu }$ can be written in terms of the BD parameter $\varpi $ as
follows 
\begin{eqnarray}
R_{\mu \nu } &=&\frac{8\pi }\phi \left[ T_{M\,\mu \nu }-\left( \frac{%
1+\varpi }{3+2\varpi }\right) g_{\mu \nu }T_{M\;\sigma }^{\;\sigma }\right]
\label{eq32} \\
&&\ +\frac \varpi {\phi ^2}\phi ;_\mu \phi ;_\nu +\frac 1\phi \phi ;_\mu
;_\nu -\frac \kappa {\phi ^2}\phi ;_\mu \phi ;_\nu  \nonumber
\end{eqnarray}
which is the same as the equivalent equation in the BD theory except with
the addition of the last term which includes the ''creation coefficient''$%
\kappa $. This expression is used in order to compare our solution with the
standard BD theory below.

\subsection{The Post-Newtonian Approximation}

In order to develop the gravitational theory consider the gravitational and
scalar fields around a static, spherically symmetric, mass embedded in a
cosmological space-time. In such an embedding the value of the scalar $\phi $
defining inertial mass is assumed to asymptotically approach a
''cosmological'' value $G_0^{\;-1}$ which holds ''at great distance'' from
any large masses. $\phi $ is determined in the inertial, Lorentz frame of
reference of the Centre of Mass using electromagnetic methods and this is
the origin of our coordinate system. 
\begin{equation}
\phi =G_0^{\;-1}\left( 1+\epsilon \right)  \label{eq33}
\end{equation}
where $G_0$ is a constant of dimension and order $G_N$, and $\epsilon $ a
scalar field defined by 
\begin{equation}
\Box \epsilon =\epsilon ;^\sigma ;_\sigma =\frac{8\pi }{3+2\varpi }%
G_0T_{\;\sigma }^\sigma  \label{eq34a}
\end{equation}
in which $\epsilon \rightarrow 0$ as $r\rightarrow \infty $ and we note $%
\varpi $ is the BD parameter $\varpi =\frac 1\lambda -\frac 32$. $T_{\mu \nu
}$ is the energy-momentum tensor of ordinary matter and energy excluding the
energy of the $\phi $ field.

The gravitational field Equation \ref{eq32} now becomes 
\begin{eqnarray}
R_{\mu \nu } &=&8\pi G_0\left( 1+\epsilon \right) ^{-1}\left[ T_{M\,\mu \nu
}-\left( \frac{1+\varpi }{3+2\varpi }\right) g_{\mu \nu }T_{M\;\sigma
}^{\;\sigma }\right]  \label{eq35} \\
&&\ +\frac \varpi {\left( 1+\epsilon \right) ^2}\epsilon ;_\mu \epsilon
;_\nu +\frac 1{\left( 1+\epsilon \right) }\epsilon ;_\mu ;_\nu -\frac \kappa
{\left( 1+\epsilon \right) ^2}\epsilon ;_\mu \epsilon ;_\nu  \nonumber
\end{eqnarray}
which again is the same as the equivalent BD equation except with the
addition of the last term which includes $\kappa $.

In the Post-Newtonian Approximation (PNA) slowly moving particles bound by
gravitational forces are considered. If $\overline{r}$ and $\overline{v}$
are typical distances and velocities of the system then the components of
the metric and the Ricci tensor are expressed in powers of the parameters $%
\frac{GM}{\overline{r}}$ and $\overline{v}^2$ and the PNA requires an
expansion of these parameters to one order beyond Newtonian mechanics.

Now $\stackrel{N}{g}_{ij}$is of the order $\overline{v}^N$, $\overline{v}%
^2\simeq \frac{GM}{\overline{r}}$ and $\stackrel{N}{R}_{\mu \nu }$is of the
order $\frac{\overline{v}^N}{\overline{r}^2}$ . Therefore in the PNA

we need to know $g_{ij}$ to order $\stackrel{2}{g}_{ij}$, $g_{i0}$ to order $%
\stackrel{3}{g}_{i0}$, and $g_{00}$ to orders $\stackrel{2}{g}_{00}$ and $%
\stackrel{4}{g}_{00}$.

The PNA formulas for the Ricci tensor are 
\begin{equation}
\stackrel{2}{R_{00}}=-\frac 12\nabla ^2\stackrel{2}{g}_{00}\text{ ,}
\label{eq36}
\end{equation}
\begin{eqnarray}
\stackrel{4}{R}_{00} &=&-\frac 12\nabla ^2\stackrel{4}{g}_{00}+\frac 12\frac{%
d^2\left( \stackrel{2}{g}_{00}\right) }{dt^2}+\frac 12\stackrel{2}{g}_{ij}%
\frac{d^2\left( \stackrel{2}{g}_{00}\right) }{dx^idx^j}  \label{eq37} \\
&&\ \ -\frac 12\left( \nabla ^2\stackrel{2}{g}_{00}\right) ^2\text{ ,} 
\nonumber
\end{eqnarray}
\begin{equation}
\stackrel{3}{R}_{0i}=-\frac 12\nabla ^2\stackrel{3}{g}_{i0}\text{ ,}
\label{eq38}
\end{equation}
and 
\begin{equation}
\stackrel{2}{R}_{ij}=-\frac 12\nabla ^2\stackrel{2}{g}_{ij}\text{ ,}
\label{eq39}
\end{equation}
where the indices $i$, $j$ run from 1 to 3, covering only the space
dimensions. From Equation \ref{eq34} $\epsilon $ has the expansion 
\begin{equation}
\epsilon =\stackrel{2}{\,\epsilon }+\stackrel{4}{\epsilon }+...  \label{eq40}
\end{equation}
where $\stackrel{N}{\epsilon }$ is of the order $\overline{v}^N$ and in
particular 
\begin{equation}
\nabla ^2\stackrel{2}{\epsilon }\,=-\frac{8\pi G_0}{3+2\varpi }\stackrel{0}{T%
}^{00}  \label{eq41}
\end{equation}
the field equations for the components of the metric, $\stackrel{2}{g}_{00}$%
, $\stackrel{4}{g}_{00}$, $\stackrel{3}{g}_{i0}$and $\stackrel{2}{g}_{ij}$%
are now given by substituting Equations \ref{eq35}, \ref{eq40} and \ref{eq41}
in Equations \ref{eq36} - \ref{eq39} as 
\begin{equation}
\nabla ^2\stackrel{2}{g}_{00}=-8\pi G_0\left( \frac{2\varpi +4}{2\varpi +3}%
\right) \stackrel{0}{T}^{00}\text{ ,}  \label{eq42}
\end{equation}
\newpage
\begin{eqnarray}
\nabla ^2\stackrel{4}{g}_{00} &=&\frac{\partial ^2\left( \stackrel{2}{g}%
_{00}\right) }{\partial t^2}+\stackrel{2}{g}_{ij}\frac{\partial ^2\left( 
\stackrel{2}{g}_{00}\right) }{\partial x^i\partial x^j}-\left( \nabla ^2%
\stackrel{2}{g}_{00}\right) ^2  \label{eq43} \\
&&\ \ +8\pi G_0\left( \frac{2\varpi +4}{2\varpi +3}\right) \stackrel{2}{%
\epsilon }\stackrel{0}{T}^{00}-8\pi G_0\left( \frac{2\varpi +2}{2\varpi +3}%
\right) \stackrel{2}{T}^{ii}  \nonumber \\
&&\ \ +16\pi G_0\stackrel{2}{g}_{00}\left( \frac{2\varpi +4}{2\varpi +3}%
\right) \stackrel{0}{T}^{00}  \nonumber \\
&&\ \ -8\pi G_0\left( \frac{2\varpi +4}{2\varpi +3}\right) \stackrel{2}{T}%
^{00}\ -2\left( \varpi -\kappa \right) \left( \frac{\partial \stackrel{2}{%
\epsilon }}{\partial t}\right) ^2  \nonumber \\
&&\ \ -2\frac{\partial ^2\stackrel{2}{\epsilon }}{\partial t^2}+2\stackrel{2%
}{\Gamma }_{\;00}^i\frac{\partial \stackrel{2}{\epsilon }}{\partial x^i}%
\text{ ,}  \nonumber
\end{eqnarray}

\begin{equation}
\nabla ^2\stackrel{3}{g}_{i0}=16\pi G_0\stackrel{1}{T}^{i0}-\frac{2\partial
^2\stackrel{2}{\epsilon }}{\partial x^i\partial t}\text{ ,}  \label{eq44}
\end{equation}
and 
\begin{equation}
\nabla ^2\stackrel{2}{g}_{ij}=-8\pi G_0\delta _{ij}\left( \frac{2\varpi +2}{%
2\varpi +3}\right) \stackrel{0}{T}^{00}-\frac{2\partial ^2\stackrel{2}{%
\epsilon }}{\partial x^i\partial x^j}\text{ .}  \label{eq45}
\end{equation}
From Equation \ref{eq42} it follows that if the usual relation between $%
\stackrel{2}{g}_{00}$ and the purely gravitational Newtonian potential $\Phi
_m$ holds by defining $\Phi _m$ as 
\begin{equation}
\nabla ^2\Phi _m=4\pi G_m\stackrel{0}{T}^{00}  \label{eq46}
\end{equation}
so normalized that 
\[
\Phi _m\left( \infty \right) =0 
\]
where $G_m$ is the metric gravitational ''constant'' associated with the
curvature of space-time measured in the limit $r\rightarrow \infty $. Then $%
G_m$ is related to $G_0$ by the relationship 
\begin{equation}
G_m=\left( \frac{2\varpi +4}{2\varpi +3}\right) G_0  \label{eq47}
\end{equation}
It is important to note that in the BD theory, where $G_m\equiv G_N$, the
definitions of $G_0$, $\phi $, and $\psi $ in Equations \ref{eq1} and \ref
{eq33} give the result 
\begin{equation}
\psi =\left( \frac{2\varpi +4}{2\varpi +3}\right)  \label{eq48}
\end{equation}
Thus $\psi $ is not necessarily unity and has to be determined in this calculation.
The relationship between $\epsilon $ and $\Phi _m$ can be derived from the
Equations \ref{eq41} and \ref{eq46} to be 
\begin{equation}
\stackrel{2}{\epsilon }\,=-\frac 1{\varpi +2}\Phi _m  \label{eq50}
\end{equation}
Note the solution for $\Phi _m$ around a spherically symmetric mass in vacuo
is given by 
\[
\Phi _m=-\frac{G_mM}r 
\]
therefore 
\begin{equation}
\stackrel{2}{\epsilon }\,=+\frac{G_mM}{\left( 2+\varpi \right) r}
\label{eq51}
\end{equation}
The field equations for $\stackrel{4}{g}_{00}$ , $\stackrel{3}{g_{i0}}$ and $%
\stackrel{2}{g}_{ij}$ are 
\begin{eqnarray}
\nabla ^2\stackrel{4}{g}_{00} &=&-2\left( \frac{\varpi +1}{\varpi +2}\right) 
\frac{\partial ^2\Phi _m}{\partial t^2}-2\stackrel{2}{g}_{ij}\frac{\partial
^2\Phi _m}{\partial x^i\partial x^j}  \label{eq52} \\
&&\ \ \ \ -2\left( \frac{2\varpi +5}{\varpi +2}\right) \left( \nabla \Phi
_m\right) ^2  \nonumber \\
&&\ \ \ -8\pi G_m\left( 4+\frac 1{\varpi +2}\right) \Phi _m\stackrel{0}{T}%
^{00}  \nonumber \\
&&\ \ \ -8\pi G_m\left( \frac{\varpi +1}{\varpi +2}\right) \stackrel{2}{T}%
^{ii}\ -8\pi G_m\stackrel{2}{T}^{00}  \nonumber \\
&&\ \ \ -2\frac{\left( \varpi -\kappa \right) }{\left( \varpi +2\right) ^2}%
\left( \frac{\partial \Phi _m}{\partial t}\right) ^2\text{ ,}  \nonumber
\end{eqnarray}
\begin{equation}
\nabla ^2\stackrel{3}{g}_{i0}=16\pi G_m\left( \frac{2\varpi +3}{2\varpi +4}%
\right) \stackrel{1}{T}^{i0}+\left( \frac 2{\varpi +2}\right) \frac{\partial
^2\Phi _m}{\partial x^i\partial t}\text{ ,}  \label{eq53}
\end{equation}
and 
\begin{equation}
\nabla ^2\stackrel{2}{g}_{ij}=-8\pi G_m\left( \frac{\varpi +1}{\varpi +2}%
\right) \delta _{\;j}^i\stackrel{0}{T}^{00}+\left( \frac 2{\varpi +2}\right) 
\frac{\partial ^2\Phi _m}{\partial x^i\partial x^j}\text{ .}  \label{eq54}
\end{equation}
The One-Body problem considers a static spherically symmetric mass. On
inspection of the above expressions for $\stackrel{4}{g}_{00}$ , $\stackrel{3%
}{g}_{i0}$ and $\stackrel{2}{g}_{ij}$the only difference between them and
their equivalents in BD [Weinberg, (1972), page 246, equations 9.9.14-9.9.16] 
is the last term in Equation \ref{eq52} in which the
coefficient $\left( \varpi -\kappa \right) $, i.e. $\omega $, replaces $%
\varpi $. However in the static case the term $\frac{d\Phi _m}{dt}$ vanishes
and so this term drops out of the solution. Therefore in the static PNA
there is no difference between this theory and BD. Hence, as with BD, the
gravitational field outside a static, spherically symmetric mass depends on $%
M$ alone but not any other property of the mass. Also in this case the
solutions of the PNA are exactly the same as with BD. Therefore the
Robertson parameters for this theory are also given by the same formulas as
in BD 
\begin{equation}
\alpha =1\text{ , }\beta =1\text{ , }\gamma =\frac{\varpi +1}{\varpi +2}%
\text{ .}  \label{eq55}
\end{equation}

\subsection{Solving for $\phi $}

The effect of allowing $T_{M\;\nu ;\mu }^{\;\mu }\neq 0$ has now to be
calculated and its effect included in the modelling of experiments of slowly
moving particles. That is the violation of the equivalence principle will
produce a force $G_\nu $ that will perturb particles, but not photons, from
their geodesic world lines. The force density is given by 
\begin{equation}
G_\nu =T_{M\;\nu ;\mu }^{\;\mu }\text{ .}  \label{eq49}
\end{equation}
In order to calculate this effect $\phi $ has to be determined to the third
order of accuracy, ($\stackrel{4}{\epsilon }$), and this is possible both in
BD and SCC. In the PNA solution to the One-Body problem the solution for $%
\phi $ obtained from Equations \ref{eq33}, \ref{eq40} and \ref{eq51} is 
\begin{equation}
\phi =G_0^{-1}\left[ 1+\frac{G_mM}{\left( 2+\varpi \right) r}+...\right]
\label{eq56}
\end{equation}
and when the metric takes the standard form of the Robertson expansion 
\begin{eqnarray}
d\tau ^2 &=&\left( 1-\frac{2G_mM}r+...\right) dt^2-\left( 1+\frac{2\gamma
G_mM}r+...\right) dr^2  \label{eq57} \\
&&\ \ +r^2d\theta ^2+r^2\sin ^2\theta d\varphi ^2\text{ ,}  \nonumber
\end{eqnarray}
where $\gamma =\frac{\left( \varpi +1\right) }{\left( \varpi +2\right) }$ .
Then, as $\Box ^2\phi =0$ in vacuo, 
\begin{equation}
\Box ^2\phi =\frac{d^2\phi }{dx^{\lambda \,\,2}}+\Gamma _{\mu \lambda }^\mu 
\frac{d\phi }{dx^{\lambda \,}}=0\text{ ,}  \label{eq58}
\end{equation}
where the affine connection $\Gamma _{\mu \lambda }^\mu $ is given by 
\begin{equation}
\Gamma _{\mu \nu }^\lambda =\frac 12g^{\lambda \rho }\left( \frac{\partial
g_{\rho \mu }}{\partial x^\nu }+\frac{\partial g_{\rho \nu }}{\partial x^\mu 
}-\frac{\partial g_{\mu \nu }}{\partial x^\rho }\right) \text{ .}
\label{eq59}
\end{equation}
As $g_{\mu \nu }$ is diagonal and $\phi =\phi \left( r\right) $, the only
non-vanishing components of the affine connection are 
\begin{eqnarray}
\Gamma _{tt}^t &=&\frac 12g^{00}\frac{dg_{00}}{dr}=+\frac{G_mM}{r^2}\left[
1+O\left( \frac{G_mM}r\right) ...\right] \text{ ,}  \label{eq60} \\
\Gamma _{rr}^r &=&\frac 12g^{rr}\frac{dg_{rr}}{dr}=-\frac{\left( \varpi
+1\right) }{\left( \varpi +2\right) }\frac{G_mM}{r^2}\left[ 1+O\left( \frac{%
G_mM}r\right) ...\right] \text{ ,}  \nonumber \\
\Gamma _{\phi r}^\phi &=&\frac 12g^{\phi \phi }\frac{dg_{\phi \phi }}{dr}%
=\frac 1r\text{, and\qquad }\Gamma _{\theta r}^\theta =\frac 12g^{\theta
\theta }\frac{dg_{\theta \theta }}{dr}=\frac 1r\text{ ,}  \nonumber
\end{eqnarray}
and Equation \ref{eq58} becomes 
\begin{equation}
\Box ^2\phi =\frac{d^2\phi }{dr^{\,2}}+\left\{ \left[ 1-\frac{\left( \varpi
+1\right) }{\left( \varpi +2\right) }\right] \frac{G_mM}{r^2}+\frac
2r+...\right\} \frac{d\phi }{dr}=0\text{ .}  \label{eq61}
\end{equation}
Integrating twice w.r.t. $r$ , and expanding the exponential with $\frac{G_mM%
}r\ll 1$ , produces a solution with two integration constants, $k_1$ and $%
k_2 $; 
\begin{equation}
\phi =k_1+\frac{k_2}r+\frac{k_2}{2\left( \varpi +2\right) }\frac{G_mM}{r^2}%
+...\text{ .}  \label{eq62}
\end{equation}
Comparing coefficients with Equation \ref{eq56} evaluates $k_1$ and $k_2$
and 
\begin{equation}
\phi =G_0^{-1}\left\{ 1+\frac{G_mM}{\left( 2+\varpi \right) r}+\frac
12\left[ \frac{G_mM}{\left( 2+\varpi \right) r}\right] ^2+...\right\} \text{
,}  \label{eq63}
\end{equation}
so, to the accuracy of the post-post Newtonian approximation, 
\begin{equation}
\phi =\phi _0\exp \left[ \frac{G_mM}{\left( 2+\varpi \right) r}\right] \text{
.}  \label{eq64}
\end{equation}
Therefore 
\begin{equation}
\frac 1\phi \frac{d\phi }{dr}=-\frac{G_mM}{\left( 2+\varpi \right) r^2}\text{
.}  \label{eq65}
\end{equation}

\subsection{The Scalar Field Acceleration}

The expression for $T_{M\;\nu ;\mu }^{\;\mu }$ for a system of n particles
of rest mass $m_n$ is given by 
\begin{equation}
T_{M\;\nu ;\mu }^{\;\mu }=G_\nu =\stackunder{n}{\sum }\delta ^3\left\{
x-x_n\left( t\right) \right\} g_{\nu \alpha }\frac{d\tau }{dt}\frac
d{dt}\left[ m_n\frac{dx^\alpha }{d\tau }\right] \text{ ,}  \label{eq66}
\end{equation}
where $\delta ^3\left\{ x-x_n\left( t\right) \right\} $ is the Dirac delta
function, $d\tau $ the proper time defined by 
\begin{equation}
d\tau ^2=-g_{\mu \nu }dx^\mu dx^\nu  \label{eq67}
\end{equation}
and $G_\nu $ is the force density.

Over the elemental volume of an individual test particle with density of
inertial rest mass $\rho \left( r,t\right) $ becomes 
\[
T_{M\;\nu ;\mu }^{\;\mu }=g_{\alpha \nu }\frac{d\tau }{dt}\frac d{dt}\left[
\rho \frac{dx^\alpha }{d\tau }\right] \text{ ,} 
\]
that is, 
\begin{equation}
T_{M\;\nu ;\mu }^{\;\mu }=g_{\alpha \nu }\left[ \frac{d\rho }{dt}\frac{%
dx^\alpha }{dt}+\rho \frac{d^2x^\alpha }{dt^2}\right] \text{ .}  \label{eq68}
\end{equation}
Substituting Equation \ref{eq2} in \ref{eq26} yields 
\begin{equation}
T_{M\;\nu ;\mu }^{\;\mu }=\frac{\kappa \lambda }2\frac{\phi ;_\nu }\phi
T_{M\;\sigma }^{\;\sigma }  \label{eq69}
\end{equation}
and using 
\begin{equation}
T_{M\;\sigma }^{\;\sigma }=3p-\rho =-\rho  \label{eq70}
\end{equation}
for a perfect fluid in the rest frame when the pressure is negligible, the
PMI solution for the non-conservation of the energy-momentum tensor, becomes 
\begin{equation}
T_{M\;\nu ;\mu }^{\;\mu }=-\frac{\kappa \lambda }2\frac{\phi ;_\nu }\phi
\rho \text{ .}  \label{eq71}
\end{equation}
The effect of the scalar field force is therefore 
\begin{equation}
g_{\alpha \nu }\left[ \frac{d\rho }{dt}\frac{dx^\alpha }{dt}+\rho \frac{%
d^2x^\alpha }{dt^2}\right] =-\frac{\kappa \lambda }2\frac{\phi ;_\nu }\phi
\rho  \label{eq72}
\end{equation}
Now consider the effect of this force on a mass particle momentarily at rest
in the frame of reference of the Centre of Mass, that is: $\frac{dx^\alpha }{%
dt}=0$. Equation \ref{eq72} becomes 
\begin{equation}
\frac{d^2x^\alpha }{dt^2}=-g^{\alpha \nu }\frac{\kappa \lambda }2\frac{\phi
;_\nu }\phi \text{ .}  \label{eq73}
\end{equation}

\subsection{Equations of Motion}

It is now possible to examine the equations of motion in this theory. At
every space-time event in an arbitrary gravitational field we can specify a
set of coordinates $\xi ^i$ in which the local description of space-time is
Minkowskian, with a Special Relativity metric $\eta _{\alpha \beta }$ and in
which a photon has an equation of motion 
\begin{equation}
\frac{d^2\xi ^\alpha }{d\sigma ^2}=0\text{ ,}  \label{eq74}
\end{equation}
\begin{equation}
0=-\eta _{\alpha \beta }\frac{d\xi ^\alpha }{d\sigma }\frac{d\xi ^\beta }{%
d\sigma }\text{ ,}  \label{eq75}
\end{equation}
where $\sigma \equiv \xi ^0$ is a suitable parameter describing the
null-geodesic. We now consider the equation of motion of a distant particle,
momentarily stationary, in the coordinate system $x^\mu $ of the frame of
reference of the Centre of Mass. Transforming coordinates into this system
the particle would also experience the scalar field acceleration described
in Equation \ref{eq73} and as the affine connection is defined by 
\[
\Gamma _{\mu \nu }^\alpha =\frac{\partial x^\alpha }{\partial \xi ^\beta }%
\frac{\partial ^2\xi ^\beta }{\partial x^\mu \partial x^\nu }\text{ ,} 
\]
then, if the pressure is negligible, 
\begin{equation}
\frac{d^2x^\alpha }{d\tau ^2}+\Gamma _{\mu \nu }^\alpha \frac{dx^\mu }{d\tau 
}\frac{dx^\nu }{d\tau }=-g^{\alpha \nu }\frac{\kappa \lambda }2\frac{\phi
;_\nu }\phi \text{ .}  \label{eq76}
\end{equation}
Therefore for a slow particle 
\begin{equation}
\frac{d^2x^\alpha }{d\tau ^2}+\Gamma _{00}^\alpha \left( \frac{dt}{d\tau }%
\right) ^2=-g^{\alpha \nu }\frac{\kappa \lambda }2\frac{\phi ;_\nu }\phi 
\text{ ,}  \label{eq77}
\end{equation}
also 
\[
\frac{d^2t}{d\tau ^2}=0\text{ ,}\qquad \text{so\qquad }\frac{dt}{d\tau }=%
\sqrt{-g_{00}^{-1}}\text{ which is a constant at }r=r_1\text{ .} 
\]
So multiplying through by $\left( \frac{d\tau }{dt}\right) ^2$ produces 
\begin{equation}
\frac{d^2x^\alpha }{dt^2}+\Gamma _{00}^\alpha =+g_{00}\,g^{\alpha \nu }\frac{%
\kappa \lambda }2\frac{\phi ;_\nu }\phi \text{ .}  \label{eq78}
\end{equation}
Now for a stationary field 
\begin{equation}
\Gamma _{00}^\alpha =-\frac 12g^{\alpha \nu }\frac{\partial g_{00}}{dx^\nu }
\label{eq79}
\end{equation}
and for a weak field, $g_{\alpha \beta }=\eta _{\alpha \beta }+h_{\alpha
\beta }$ , where $\mid h_{\alpha \beta }\mid \ll 1$ and $\eta _{\alpha \beta
}$. The resulting affine connection, linearized in the metric perturbation $%
h_{\alpha \beta }$, becomes in the spherically symmetric case, (as in GR) 
\begin{equation}
\Gamma _{00}^r=-\frac 12\eta _{}^{rr}\frac{dh_{00}}{dr}  \label{eq80}
\end{equation}
$\label{eq97f}$so the only non zero component of Equation \ref{eq78} is 
\begin{equation}
\frac{d^2r}{dt^2}=\frac 12\frac{dh_{00}}{dr}+g_{00}\,g^{rr}\frac{\kappa
\lambda }2\frac 1\phi \frac{d\phi }{dr}\text{ .}  \label{eq81}
\end{equation}
The general standard form of the metric in both the BD and SCC theories is 
\begin{eqnarray}
d\tau ^2 &=&\left[ 1-\frac{2G_mM}r+2\left( 1-\gamma \right) \left( \frac{G_mM%
}r\right) ^2+...\right] dt^2  \label{eq82} \\
&&\ \ \ \ -\left( 1+\frac{2\gamma G_mM}r+...\right) dr^2+r^2d\theta
^2+r^2\sin ^2\theta d\varphi ^2\text{ ,}  \nonumber
\end{eqnarray}
where in both theories $\gamma =\frac{\left( 2-\lambda \right) }{\left(
2+\lambda \right) }$ , therefore 
\begin{equation}
g_{00}=-\left[ 1-\frac{2G_mM}r+\frac{4\lambda }{\left( 2+\lambda \right) }%
\left( \frac{G_mM}r\right) ^2+...\right] \text{ ,}  \label{eq83}
\end{equation}
\begin{equation}
h_{00}=\frac{2G_mM}r-\frac{4\lambda }{\left( 2+\lambda \right) }\left( \frac{%
G_mM}r\right) ^2+...  \label{eq84}
\end{equation}
and 
\begin{equation}
g_{rr}=1+\frac{2\left( 2-\lambda \right) G_mM}{\left( 2+\lambda \right) r}%
+...\text{ .}  \label{eq85}
\end{equation}
Substituting Equations \ref{eq65}, \ref{eq83}, \ref{eq84} and \ref{eq85} in
Equation \ref{eq81} yields 
\begin{eqnarray}
\frac{d^2r}{dt^2} &=&-\left[ 1-\frac{\kappa \lambda ^2}{\left( 2+\lambda
\right) }\right] \frac{G_mM}{r^2}  \label{eq86} \\
&&\ \ \ \ \ \ \ \ +\left\{ \frac{4\lambda \left( 2+\lambda -2\kappa \lambda
\right) }{\left( 2+\lambda \right) ^2}\right\} \frac{\left( G_mM\right) ^2}{%
r^3}+...  \nonumber
\end{eqnarray}
and therefore to first order the total acceleration experienced by a
particle is 
\begin{equation}
\frac{d^2r}{dt^2}=-\left[ 1-\frac{\kappa \lambda ^2}{2+\lambda }\right] 
\frac{G_mM}{r^2}+...\text{ .}  \label{eq87}
\end{equation}
However Newtonian gravitational theory defines $G_{N\text{ }}$ by 
\begin{equation}
\frac{d^2r}{dt^2}=-\frac{G_NM}{r^2}\text{ ,}  \label{eq88}
\end{equation}
therefore the effect of violating the equivalence principle in accordance
with the PMI is that every mass experiences an extra acceleration similar to
Newtonian gravitation and which therefore is confused with it. According to
SCC the Newtonian gravitational constant $G_N$, as measured in a Cavendish
type experiment, is a compilation of the effect of the curvature of space
time, with its corresponding $G_m$, and the action of the scalar field. The
total Newtonian acceleration experienced by a mass particle is therefore 
\begin{equation}
G_N=\left[ 1-\frac{\kappa \lambda ^2}{\left( 2+\lambda \right) }\right] G_m%
\text{ .}  \label{eq89}
\end{equation}
Note that $G_N$ and $G_m$ refer to the total ''gravitational'' accelerations
experienced in physical experiments by atomic particles and photons
respectively.

\section{Incorporating the Local Conservation of Energy in BD}

\subsection{The Relationship Between $\phi $ And $m$.}

Consider the general Gauss Divergence theorem applied to the gradient of the
Newtonian potential $\Phi _N$%
\[
\stackunder{V}{\iiint }\nabla \Theta \circ dV=\stackunder{S}{\iint }\Theta
\circ dS\text{ ,}\qquad 
\]
\[
\text{put }\underline{\Theta }=\underline{\nabla }\Phi _N\quad \text{and
define }\Phi _N\text{ by\quad }\nabla ^2\Phi _N=4\pi G_N\rho \quad \text{%
with\quad }\stackunder{r\rightarrow \infty }{Lim}\Phi _N\left( r\right) =0%
\text{ ,} 
\]
\begin{equation}
\stackunder{V}{\iiint }\nabla ^2\Phi _N\,dV=\stackunder{S}{\iint }\underline{%
\nabla }\Phi _N.\underline{dS}\text{ .}  \label{eq113}
\end{equation}
In the spherically symmetric One Body case the volume integral on the left
hand side is simply $4\pi GM$ where $M$ is the remote determination of the
total mass of the central body radius $R$. Consider several concentric
external spheres of radius $r_1$, $r_2$ etc. $\geq R$ centered on the mass $%
M $ . As the contributions from the vacuum are zero the volume integrals
over each sphere are equal. 
\begin{equation}
\stackunder{V_1}{\iiint }\nabla ^2\Phi _N.dV=\stackunder{V_2}{\iiint }\nabla
^2\Phi _N.dV=4\pi GM\text{ .}  \label{eq94a}
\end{equation}
Therefore observers on the surface of each sphere will have different
determinations of the central mass, which will vary $M\propto m_i^{-1}$in
the JF(E), that is when comparing $M$ to their locally determined atomic
masses $m_i$ by observing the red shift of photons that are emitted from the central 
mass with invariant energy. As $GM$ is constant for all $r\geq R$ they
will conclude 
\[
G\left( r\right) \propto M^{-1}\left( r\right) \propto m_i\left( r\right) 
\text{ .} 
\]
But $G(r)=\frac \psi \phi $ therefore consistency demands 
\begin{equation}
\phi \left( r\right) \propto m_i\left( r\right) ^{-1}\text{ .}  \label{eq114}
\end{equation}

\subsection{The Gravitational Red-Shift of Light}

In order to examine the measurement problem in both the EF and the JF(E) the
gravitational red shift of light is now considered. This analysis depends on
the assumption that if no work is done on, or by, a projectile while in free
fall then its energy $E$ , $P^0$ , is conserved when measured in a specific
frame of reference, that of the CoM of the system.\textit{\ }In a gedanken,
'thought', experiment, construct a laboratory at the co-moving centroid, the
CoM, of the system. Connect it to the outside world by a radial tube through
which identical test masses and photons may be projected \underline{in vacuo}%
. Launch such projectiles, with rest masses, $m_0$, at the CoM at various
velocities to reach maximum altitudes $r_i$ where $r_i$ varies from $R$, the
radius of the central mass, out to infinity. The mass of the projectile $%
m_c\left( r\right) $ , the 'coordinate' mass, is in general a function of
altitude measured in the CoM frame of reference.

First consider such a photon emitted by one atom at altitude $x_2$ and
absorbed by another at an altitude $x_1$. The emission and absorption
frequencies of the photon, $\nu \left( x_2\right) $ and $\nu \left(
x_1\right) $, are determined by comparing the arrival times of two adjacent
wave fronts emitted from one point in a gravitational field at $\left(
x_2\right) $ and received at another at $\left( x_1\right) $. The standard
time dilation relationship is thereby derived 
\begin{equation}
\frac{\nu \left( x_2\right) }{\nu \left( x_1\right) }=\left[ \frac{%
-g_{00}\left( x_2\right) }{-g_{00}\left( x_1\right) }\right] ^{\frac 12}%
\text{ .}  \label{eq139}
\end{equation}
Hence substituting $x_2=r$ and $x_1=\infty $ in Equation \ref{eq139}, where $%
g_{00}\left( x_1\right) =-1$, and writing $\nu \left( \infty \right) $ as $%
\nu _0$, yields the standard (GR) gravitational red shift relationship 
\begin{equation}
\nu \left( r\right) =\nu _0\left[ -g_{00}\left( r\right) \right] ^{\frac 12}%
\text{ .}  \label{eq140}
\end{equation}
Where the observer is at infinite altitude receiving a photon emitted at
altitude $r$.

Now consider the various projectiles. With the standard definition of proper
time $\tau $ from the metric 
\begin{equation}
d\tau ^2=-g_{\mu \nu }dx^\mu dx^\nu \text{ .}  \label{eq141}
\end{equation}
The 4-momentum vector of the projectile is defined 
\begin{equation}
P^\mu =m_c\frac{dx^\mu }{d\tau }\text{ .}  \label{eq142}
\end{equation}
The time component of 4-momentum $P^\mu $ is the total 'relative' energy $E$
and the space components form the 'relative' 3-momentum $\underline{p}$ .
Now from Equation \ref{eq141} 
\begin{equation}
\frac{d\tau ^2}{dt^2}=-g_{00}-2g_{i0}v^i-v^2\text{ ,}  \label{eq144}
\end{equation}
\begin{equation}
\text{where }v^i=\frac{dx^i}{dt}\text{ and }v^2=g_{ij}\frac{dx^i}{dt}\frac{%
dx^j}{dt}\text{ .}  \label{eq145}
\end{equation}
Therefore in a spherically symmetric, non-rotating, metric with $g_{i0}=0$, 
\begin{equation}
-g_{00}E^2=m_c^2+\underline{p}^2\text{ .}  \label{eq148}
\end{equation}
This is the spherically symmetric curved space-time equivalent to the SR
identity 
\begin{equation}
E^2=m_c^2+\underline{p}^2\text{ .}  \label{eq149}
\end{equation}

Now consider two of the projectiles as they momentarily reach their
respective apocentres at maximum altitude $r$, and $r+\delta r$. As they are
momentarily stationary in the CoM frame $\underline{p}=0$ . The difference
between the two adjacent projectiles at their apocentres is that one has a
total energy and rest mass of $E\left( r\right) $, and $m_c\left( r\right) $%
, and the other $E\left( r+\delta r\right) $, and $m_c\left( r+\delta
r\right) $ . Expanding for small $\delta r$, and where a prime ($^{\prime }$%
) means $\frac d{dr}$, in the limit $\delta r\rightarrow 0$ we obtain 
\begin{equation}
\frac 12\frac{\left[ -g_{00}^{\prime }\left( r\right) \right] }{\left[
-g_{00}\left( r\right) \right] }+\frac{E^{\prime }\left( r\right) }{E\left(
r\right) }=\frac{m_c^{\prime }\left( r\right) }{m_c\left( r\right) }\text{ .}
\label{eq153}
\end{equation}
Here two identical projectiles separated by an infinitesimal increase in
altitude are compared. The only difference between them is the infinitesimal
energy $\delta E$ required to raise such a projectile from $r$ to $r+\delta r
$. Although the Newtonian potential $\Phi _N\left( r\right) $ is defined by 
\begin{equation}
\nabla ^2\Phi _N\left( r\right) =4\pi G_N\stackrel{0}{T}^{00}=4\pi G_N\rho 
\label{eq154}
\end{equation}
which is normalized, 
\[
\Phi _N\left( \infty \right) =0\text{ ,}
\]
it is actually measured in a Cavendish type laboratory experiment by the
force vector acting on a body, which is given by 
\begin{equation}
\mathbf{F}=-m_p\mathbf{\nabla }\Phi _N\left( r\right) \text{ .}
\label{eq155}
\end{equation}
Then if the mass $m\left( r\right) $ is raised a height $\delta \mathbf{r}$
against this force, the infinitesimal energy $\delta E$ required is 
\begin{equation}
\delta E=-\mathbf{F}\circ \delta \mathbf{r}=m_p\left( r\right) \mathbf{%
\nabla }\Phi _N\left( r\right) \circ \delta \mathbf{r}\text{ .}
\label{eq156}
\end{equation}
That is in the radial case 
\begin{equation}
\delta E=m_p\left( r\right) \Phi _N^{\prime }\delta r\text{ ,}  \label{eq157}
\end{equation}
where $m_p\left( r\right) $ is that physical mass entering into the
Newtonian gravitational equation. Define such physical mass, momentarily at
rest, as 
\begin{equation}
m_p\left( r\right) =E\left( r\right) \text{ ,}  \label{eq158}
\end{equation}
so that the total ''relative'' energy at an altitude $r$ is its rest mass at
that altitude, measured in the CoM frame of reference. In the limit $\delta
r\rightarrow 0$ Equation \ref{eq157} becomes 
\begin{equation}
\frac{E^{\prime }(r)}{E\left( r\right) }=\Phi _N^{\prime }\left( r\right) 
\text{ ,}  \label{eq159}
\end{equation}
which when substituted in Equation \ref{eq153} yields 
\begin{equation}
\frac 12\frac{\left[ -g_{00}^{\prime }\left( r\right) \right] }{\left[
-g_{00}\left( r\right) \right] }+\Phi _N^{\prime }\left( r\right) =\frac{%
m_c^{\prime }\left( r\right) }{m_c\left( r\right) }\text{ . }  \label{eq160}
\end{equation}
This integrates directly, 
\begin{equation}
\frac 12\ln \left[ -g_{00}\left( r\right) \right] +\Phi _N\left( r\right)
=\ln \left[ m\left( r\right) \right] +k  \label{eq161}
\end{equation}
where $k$ is determined in the limit $r\rightarrow \infty $, $g_{00}\left(
r\right) \rightarrow -1$, $\Phi _N\left( r\right) \rightarrow 0$ and $%
m\left( r\right) \rightarrow m_0$. The rest mass, $m\left( r\right) $, of a
projectile at altitude $r$, evaluated in the co-moving CoM frame is
therefore given by 
\begin{equation}
m_c\left( r\right) =m_0\exp \left[ \Phi _N\left( r\right) \right] \left[
-g_{00}\left( r\right) \right] ^{\frac 12}\text{ .}  \label{eq163}
\end{equation}
This is the value, $m_c\left( r\right) $, given by an observer at infinite
altitude, where Special Relativity and a ground state solution to the theory
are recovered, with well defined particle rest mass $m_0$, 'looking down' to
a similar particle at an altitude $r$. From this expression it is obvious
that with our assumption of the conservation of energy, $P^0$ , in the CoM
frame gravitational time dilation, the factor $\left[ -g_{00}\left( r\right)
\right] ^{\frac 12}$, applies to massive particles as well as to photons. As
physical experiments measuring the frequency of a photon compare its energy
with the mass of the atom it interacts with, it is necessary to compare the
masses (defined by Equation \ref{eq163}) of two atoms at altitude, $r$ and $%
\infty $, with the energy (given by Equation \ref{eq140}) of a ''reference''
photon transmitted between them. This yields the physical rest mass $%
m_p\left( r\right) $ as a function of altitude 
\begin{equation}
\frac{m_p\left( r\right) }{\nu \left( r\right) }=\frac{m_0}{\nu _0}\exp
\left[ \Phi _N\left( r\right) \right] \text{ .}  \label{eq164}
\end{equation}

Equation \ref{eq164} is a result relating observable quantities, but how is
it to be interpreted? In other words how are mass and frequency to be
measured in any particular frame? In the GR EF (and BD JF) the physical rest
mass of the atom is defined to be constant, hence prescribing ($\widetilde{x}%
^\mu $), with $m_p\left( \widetilde{r}\right) =m_0$. In this case Equation 
\ref{eq164} becomes 
\begin{equation}
\nu \left( \widetilde{r}\right) =\nu _0\left( 1-\widetilde{\Phi }_N\left( 
\widetilde{r}\right) +...\right) \text{ .}  \label{eq165}
\end{equation}
Hence photons transmitted out of a gravitational potential well are said to
exhibit a red shift which is equal to the dimensionless Newtonian potential $%
\widetilde{\Phi }_N$, and equal in GR, ''coincidentally'', to the time
dilation effect, the factor$\left[ -\widetilde{g}_{00}\left( \widetilde{r}%
\right) \right] ^{\frac 12}$. That is, compared to reference atoms they
mysteriously appear to lose (potential) energy.

However in the SCC JF(E) rest mass is given by the expression Equation \ref
{eq24}, consequently a comparison of Equation \ref{eq164} with the equation
for rest mass in this frame yields 
\begin{equation}
\nu \left( r\right) =\nu _0\text{ .}  \label{eq168}
\end{equation}

Therefore in the SCC JF(E), in which energy is locally conserved,
gravitational red shift is interpreted not as a loss of potential energy by
the photon but rather as a gain of potential energy by the apparatus
measuring it. It is important to note that in this frame the frequency, and
hence wavelength and energy, of a free photon is invariant, even when
transversing space-time with curvature.

On the other hand, as experiments using physical apparatus refer
measurements of energy and mass to the mass of the atoms of which they are
composed, such observations interpret rest masses to be constant by
definition. In SCC such experiments are conducted in its EF in which it is
Equation \ref{eq165} that describes gravitational red shift .

Using either frame the gravitational red shift prediction in SCC is in
agreement with GR and all observations to date .

Time is the fundamental measurement in both frames, measured by the
frequency of a reference photon in the JF(E) and by the Bohr frequency of an
atom in the EF. By definition the speed of light is invariant in both.

\subsection{At the Centre of Mass}

Consider the origin of our coordinate system in the static, spherically symmetric, case, 
which is the centre of mass of the system. In Relativity theory the centroid 
of an isolated system with energy-momentum tensor $T^{\mu \nu }$ and total
4-momentum $P^\alpha $, when observed by an observer\textit{\ O }with a
4-velocity $U^\alpha $ at his Lorentz time $x^0=t$ and in his own Lorentz
frame, is defined by 
\begin{equation}
X_{\;\mu }^j\left( t\right) =\left( \frac 1{P^0}\right) \stackunder{x_0=t}{%
\int }x^jT^{00}d^3x  \label{eq97}
\end{equation}
and the co-moving centroid associated with the rest frame of the system is
defined to be its Centre of Mass (CoM). At the CoM the resultant of all
gravitational forces vanishes hence so does $\underline{\nabla }\Phi _N$.
Furthermore $\phi =\phi \left( \Phi _{N\text{ }}\right) $, therefore with $%
\underline{\nabla }\Phi _N=0$ at the CoM, 
\begin{equation}
\text{ }\nabla \phi =0\text{ .}  \label{eq98}
\end{equation}
As $\phi \left( x_\nu \right) $ is static and not dependent on time we have
for all four $\nu $%
\begin{equation}
\nabla _\nu \phi =0\text{ ,}  \label{eq99}
\end{equation}
thus at the CoM, by Equation \ref{eq26}, the PMI yields 
\begin{equation}
\nabla _\mu T_{M\;\nu }^{\;\mu }=\frac 1{8\pi \lambda }\frac 1\phi \nabla
_\nu \phi \Box \phi =0\text{ .}  \label{eq100}
\end{equation}

Hence the unique location of the centre of mass of the system, where the
energy-momentum tensor of matter is conserved with respect to covariant
differentiation, can be considered a 'proper laboratory'. Here the theory
admits a ground state solution, $g_{\mu \nu }\rightarrow \eta _{\mu \nu }$
and $\nabla _\mu \phi =0$ , here the equivalence principle holds, even for a
massive particle, and here a free falling physical clock, remaining at rest,
records proper time. Distances can be measured by timing the echo of light
rays (radar) using that clock and the metric properly defined. Also Special
Relativity is recovered as here the metric is Minkowskian and standards of
mass, length, time and the physical constants defined for atoms, together
with potential energy, retain their classical meaning. Such a standard
defined atom emits a 'reference' photon, which in the JF(E) is transmitted
across space-time with invariant energy and frequency.

In SCC the CoM preferred frame of reference is selected if and only if
energy is to be locally conserved, otherwise the equations are manifestly
covariant.

Consider again the spherical shells around a central spherical mass $M$ 
Integrating the surface integral on the right hand side of
Equation \ref{eq113} over the sphere at constant $r$ gives simply $4\pi
r^2\nabla \Phi _N\left( r\right) $, (In the standard general form of the
metric the surface area of a sphere is $4\pi r^2$. This fact is used both
here and below). Now the Newtonian potential is defined by a measurement of
acceleration in a local experiment 
\[
\frac{d^2\underline{r}}{dt^2}=-\underline{\nabla }\Phi _N\left( r\right) 
\text{ .}
\]
An observer at the Centre of Mass of the system, in the 'proper laboratory'
where $M$ is constant, would conlude from Equations \ref{eq113} and \ref
{eq94a} that 
\begin{equation}
\frac{d^2r}{dt^2}=-\frac \psi \phi \frac M{r^2}\text{ .}  \label{eq115}
\end{equation}

Reiterating, in both BD and SCC the inertial masses of fundamental particles 
$m_i$ are dependent on their interaction with the scalar field $\phi $.
However the difference is that whereas in BD a coordinate system is used in
which $m_i$ is constant and it is $G$ which is observed to vary, in the One
Body Problem of SCC, in which $m_i\equiv m_i(r)$ and $\phi \equiv \phi (r)$,
it is the rest mass $m_i$, as measured by electromagnetic radiation from the
proper laboratory, which varies, and $G_N$, as measured by atomic apparatus
in a 'local laboratory', which is invariant.

\subsection{Evaluating the Parameters in the Field Equations}

The parameters $\lambda $, $\kappa $, and $\psi $, will now be
calculated. There are two methods of calculating the combined gravitational
and scalar field acceleration, one derived from the equations of motion:
Equation \ref{eq86} and the other derived from the definition of the
Newtonian potential applied to Gauss Divergence theorem: Equation \ref{eq115}%
. Consistency between these two methods places constraints on the three
parameters. Using Equations \ref{eq47} and \ref{eq27a} the relationship
between $G_m$ and $G_0$ may be written 
\begin{equation}
G_m=\left( \frac{2+\lambda }2\right) G_0\text{ ,}  \label{eq116}
\end{equation}
and using this to substitute for $G_0$ in Equation \ref{eq86} the combined
gravitational and scalar field acceleration of a free falling massive body
is given by 
\begin{equation}
\frac{d^2r}{dt^2}=-\left\{ \frac 12\left( 2+\lambda -\kappa \lambda
^2\right) -\left( 2+\lambda -2\kappa \lambda \right) \frac{\lambda G_0M}%
r+...\right\} \frac{G_0M}{r^2}\text{ .}  \label{eq117}
\end{equation}
But we also have an expression for this combined acceleration from Equation 
\ref{eq115} together with the solution for $\phi $ in Equation \ref{eq64},
expanded for small $\frac{G_mM}r$. Using Equation \ref{eq116} this becomes 
\begin{equation}
\frac{d^2r}{dt^2}=-\psi \left[ 1-\frac{\lambda G_0M}r+...\right] \frac{G_0M}{%
r^2}\text{ .}  \label{eq118}
\end{equation}
Comparing coefficients between equations \ref{eq117} and \ref{eq118} sets
two conditions on $\lambda $ , $\kappa $, and $\psi $. Consistency between
the coefficients of $\frac{G_0M}{r^2}$ requires 
\begin{equation}
2+\lambda -\kappa \lambda ^2=2\psi  \label{eq119}
\end{equation}
and consistency between the coefficients of $\frac{\left( G_0M\right) ^2}{r^3%
}$ requires 
\begin{equation}
2+\lambda -2\kappa \lambda =\psi \text{ .}  \label{eq120}
\end{equation}

Furthermore we have two solutions for $\phi $; one from the field Equation 
\ref{eq2}, and the other from the local conservation of energy. The solution
from Equation \ref{eq64} is 
\begin{equation}
\phi =\phi _0\exp \left[ \frac{2\lambda }{2+\lambda -\kappa \lambda ^2}\frac{%
G_NM}r\right]   \label{eq121}
\end{equation}
the solution from Equations \ref{eq24} and \ref{eq114} is 
\begin{equation}
\phi =\phi _0\exp \left[ \frac{G_NM}r\right] \text{ ,}  \label{eq122}
\end{equation}
so consistency between Equations \ref{eq121} and \ref{eq122} sets a third
condition on the three parameters 
\begin{equation}
2-\lambda -\kappa \lambda ^2=0\text{ .}  \label{eq123}
\end{equation}
There are three simultaneous equations \ref{eq119}, \ref{eq120} and \ref
{eq123} for $\psi $, $\lambda $ and $\kappa $. Their unique solution is 
\begin{equation}
\psi =1\text{ ,}\qquad \lambda =1\qquad \text{and\qquad }\kappa =1\text{ .}
\label{eq124}
\end{equation}
Furthermore Equations \ref{eq89} and \ref{eq116} give the result 
\begin{equation}
G_N=\frac 12\left( 2+\lambda -\kappa \lambda ^2\right) G_0=G_0=\stackunder{%
r\rightarrow \infty }{\quad Lim}\frac 1{\phi \left( r\right) }\text{ .}
\label{eq125}
\end{equation}

Thus $G_N$ is the proper value of $\phi ^{-1}$ as measured by atomic
apparatus at infinity, and will be that value determined by physical
apparatus in ''Cavendish'' type experiments elsewhere.

\section{Conclusions}

In conclusion, by following through carefully the consequences of introducing the 
principles of mutual interaction and the local conservation of energy we have determined 
that the three parameters introduced into the equations: $\lambda $, $\kappa $ and $\psi $ are 
all unity. The values of $\lambda $ and $\kappa $ yield the following standard
formulae: from Equations \ref{eq27a} and \ref{eq27} 
\begin{equation}
\varpi =\frac 1\lambda -\frac 32=-\frac 12\text{ , \quad }  \label{eq126}
\end{equation}
and finally 
\begin{equation}
\omega =\frac 1\lambda -\frac 32-\kappa =-\frac 32\text{ ,}  \label{eq127}
\end{equation}
hence the value $\omega =-\frac 32$ required in order to make the EF of the
theory canonical GR is that value determined by consistency between Mach's
Principle and the Local Conservation of Energy in SCC. 

With this value of $\omega =-\frac 32$ the field equations of the theory become:

The scalar field equation 
\begin{equation}
\Box \phi =4\pi T_M^{\;}\text{ ,}  \label{eq128}
\end{equation}

the gravitational field equation

\begin{eqnarray}
R_{\mu \nu }-\frac 12g_{\mu \nu }R &=&\frac{8\pi }\phi T_{M\mu \nu }-\frac
3{2\phi ^2}\left( \nabla _\mu \phi \nabla _\nu \phi -\frac 12g_{\mu \nu
}g^{\alpha \beta }\nabla _\alpha \phi \nabla _\beta \phi \right)
\label{eq129} \\
&&\ \ +\frac 1\phi \left( \nabla _\mu \nabla _\nu \phi -g_{\mu \nu }\Box
\phi \right) \text{ ,}  \nonumber
\end{eqnarray}

and the creation equation, 

\begin{equation}
\nabla _\mu T_{M\;\nu }^{\;\mu }=\frac 1{8\pi }\frac 1\phi \nabla _\nu \phi
\Box \phi  \label{eq130}
\end{equation}

These field equations give rise to the cosmological and experimental features 
of the theory. The two conformal frames of the theory are both physical [see (Barber, 2002a)] 
one, the JF(E), conserves mass-energy and the other, the EF, conserves 
energy-momentum. Photons follow the geodesics of the JF, which determine cosmological 
evolution and particles follow the geodesics of GR, which determine all solar system 
experiments to date (February 2003). Those three definitive experiments [see (Barber, 2003)] 
that observe directly the curvature of space-time, or the false vacuum required by curvature, 
are the only ones able to distinguish between the two theories and are able to falsify one, 
or both, of them. The geodetic experiment of 'Gravity Probe B' is awaited with anticipation.

\subsection{Ackowlegement}

I wish to acknowledge my debt to Steven Weinberg as his textbook [Weinberg,
(1972)] has been my guide. I have used and adapted his methods in the calculations 
incorporating the PMI, although I have used a sign convention that has reversed 
the sign of $R_{\mu \nu }$ and $R$.

\end{document}